%% file: OpenGemm.tex
\documentclass[sigconf]{acmart}

\AtBeginDocument{%
  \providecommand\BibTeX{{%
    \normalfont B\kern-0.5em{\scshape i\kern-0.25em b}\kern-0.8em\TeX}}}

\usepackage{tikz}
\definecolor{RedOrange}{RGB}{255,69,0} 
\definecolor{Cerulean}{RGB}{0,123,167}
\newcommand{\idb}[1]{\ding{\numexpr181 + #1}}

\newcommand{\idblue}[1]{\color{Cerulean}\ding{\numexpr181 + #1}\color{black}}
\newcommand{\idr}[1]{\color{RedOrange}\ding{\numexpr181 + #1}\color{black}}

\usepackage{graphicx}
\usepackage{threeparttable}
\usepackage{multirow}
\usepackage{graphicx}
\usepackage{caption}
\usepackage{hhline}
\usepackage{enumitem}
\usepackage{listings}
\usepackage{pifont}
\usepackage{float}
\usepackage{colortbl}
\usepackage{tikz}
\usepackage{booktabs} 
\usepackage{tablefootnote}
\usepackage{afterpage}
\usepackage{subcaption}
\usepackage{pifont}
\usepackage{newtxmath}

\AtBeginDocument{%
  \providecommand\BibTeX{{%
    Bib\TeX}}}

\acmConference[ASP-DAC 2025]{OpenGeMM: A Highly-Efficient GeMM Accelerator Generator with Lightweight RISC-V Control and Tight Memory Coupling}{Jan. 20--23,
  2025}{Tokyo, Japan}
\copyrightyear{2025}
\acmYear{2025}
\setcopyright{acmlicensed}\acmConference[ASPDAC '25]{30th Asia and South Pacific Design Automation Conference}{January 20--23, 2025}{Tokyo, Japan}
\acmBooktitle{30th Asia and South Pacific Design Automation Conference (ASPDAC '25), January 20--23, 2025, Tokyo, Japan}
\acmDOI{10.1145/3658617.3697652}
\acmISBN{979-8-4007-0635-6/25/01}

\begin{document}
\pagestyle{empty}

\title{OpenGeMM: A High-Utilization GeMM Accelerator Generator with Lightweight RISC-V Control and Tight Memory Coupling}


\author{%
  Xiaoling Yi$^{1}$,
  Ryan Antonio$^{1}$,
  Joren Dumoulin$^{1}$,
  Jiacong Sun$^{1}$,
  Josse Van Delm$^{1}$,
  Guilherme Paim$^{1,2}$,
  Marian Verhelst$^{1}$
}
\affiliation{%
  \institution{$^{1}$MICAS-ESAT, KU Leuven}
  \city{Leuven}
  \country{Belgium}
}
\affiliation{%
  \institution{$^{2}$INESC-ID, Instituto Superior Técnico, Universidade de Lisboa}
  \city{Lisbon}
  \country{Portugal}
}








\input{content/abstract}
\keywords{Matrix Multiplication, GeMM Accelerator, Hardware Generators, RISC-V, Tight Memory Coupling, Open Source.}
\maketitle

\input{content/intro}
\input{content/gemm_arch}
\input{content/system_arch}
\input{content/results}
\input{content/conclusion}

\bibliography{reference}

\end{document}

%% file: content/abstract.tex
\begin{abstract}

Deep neural networks (DNNs) face significant challenges when deployed on resource-constrained extreme edge devices due to their computational and data-intensive nature.
While standalone accelerators tailored for specific application scenarios suffer from inflexible control and limited programmability, generic hardware acceleration platforms coupled with RISC-V CPUs can enable high reusability and flexibility, yet typically at the expense of system-level efficiency and low utilization. 

To fill this gap, we propose \textit{OpenGeMM}, an open-source acceleration platform, jointly demonstrating high efficiency and utilization, as well as ease of configurability and programmability. 
\textit{OpenGeMM} encompasses a parameterized Chisel-coded GeMM accelerator, a lightweight RISC-V processor, 
and a tightly coupled multi-banked scratchpad memory.
The GeMM core utilization and system efficiency are boosted through three mechanisms: configuration pre-loading, input pre-fetching with output buffering, and programmable strided memory access.
Experimental results show that \textit{OpenGeMM} can consistently achieve hardware utilization ranging from $81.89\%$ to $99.34\%$ across diverse CNN and Transformer workloads.
Compared to the SotA open-source Gemmini accelerator, \textit{OpenGeMM} demonstrates a $3.58 \times$ to $16.40 \times$ speedup on normalized throughput across a wide variety of GeMM workloads, while achieving $4.68$ TOPS/W system efficiency.

\end{abstract}

%% file: content/intro.tex
\section{Introduction}
DNN models have been rapidly integrated into various aspects of our society, bringing a blossom of novel applications.
However, they also bring a voracious demand for ever more computing power, presenting significant challenges for efficient execution. 
This problem is particularly severe when deploying DNNs at the edge, such as in-vehicle and wearable devices, where stringent power and area constraints exist~\cite{schizas2022tinyml}. 

A variety of domain-specific DNN accelerators have emerged in recent years, with remarkable performance and energy efficiency \cite{silvano2023survey, chen2020survey}. However, these accelerators are typically tailored to specific workloads~\cite{adiono2023fast, goetschalckx2021depfin, ham20203}, hindering their reusability across diverse applications. 
Furthermore, they are often equipped with dedicated control and data interfaces, posing substantial challenges to smoothly integrating them into standard SoCs.
For instance, deploying the NVDLA accelerator~\cite{sijstermans2018nvidia} requires additional wrappers and a custom compiler for signal translation when communicating to a host system~\cite{feng2019implementation, gonzalez2020chipyard}.

\begin{figure}[t]
    \centering
    \includegraphics[width=.9\columnwidth]{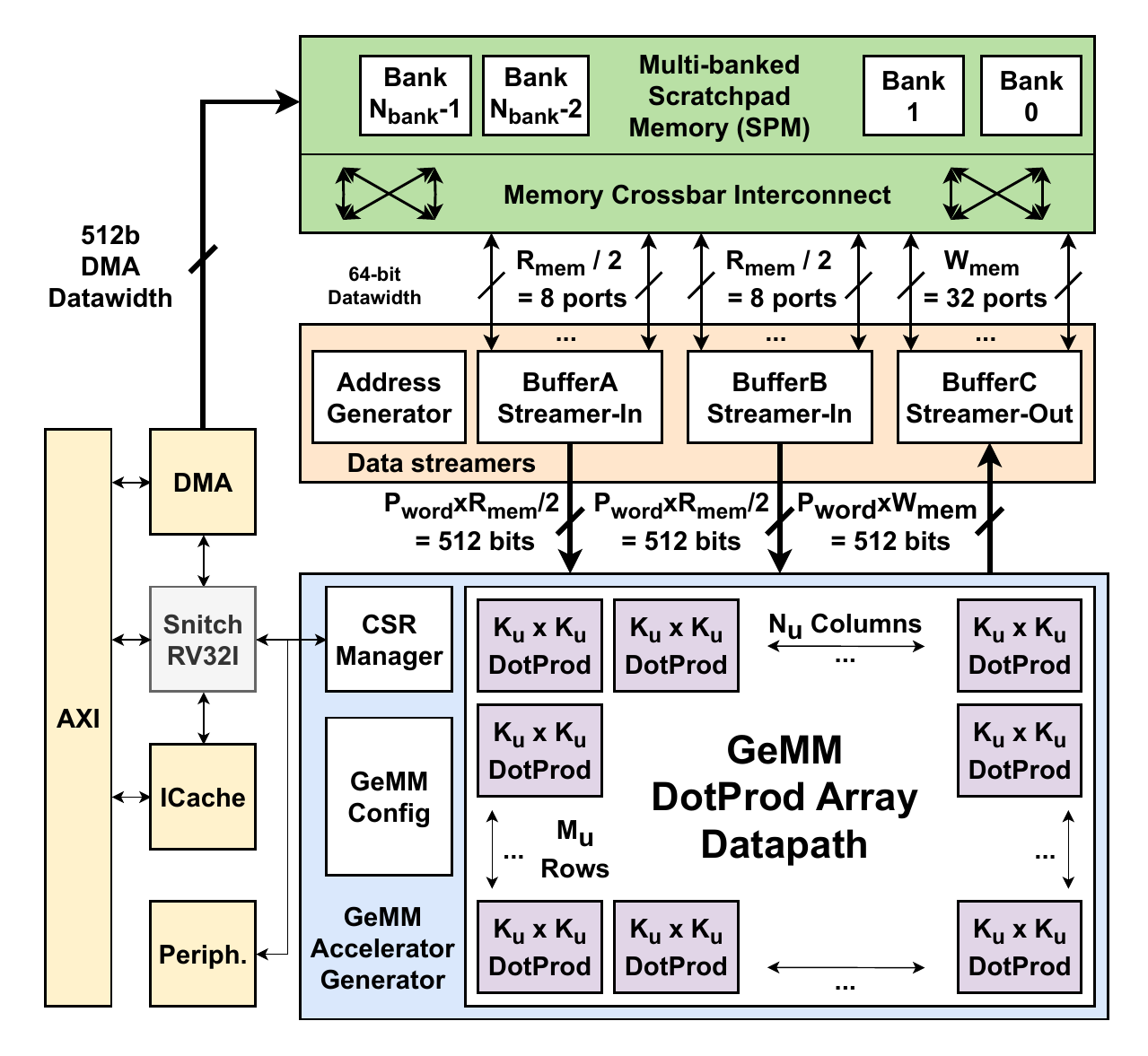}
    \vspace{-0.6cm}
    \centering
    \caption{\textit{OpenGeMM} platform overview. 
    }
    \vspace{-0.6cm}
    \label{fig:sys_arch}
\end{figure}

Alternatively, hardware acceleration platforms are proposed that include flexible DNN accelerators capable of targeting general computation kernels like general matrix multiplication (GeMM). These platforms integrate with a RISC-V host CPU, aiming to enhance hardware programmability and reusability \cite{waterman2015risc, conti2024marsellus, genc2021gemmini, park2022conna, conti2018xnor, tortorella2023redmule}.
While this solution appears promising, existing platforms face two important challenges: 1.) 
It is difficult to keep the control overhead negligible, in order to not undermine the efficiency benefits of the accelerators.
For example, the well-known Gemmini platform~\cite{genc2021gemmini} employs a hardware generator design methodology and standard RISC-V control interface \cite{savas2018designing}, facilitating rapid design-time flexibility and seamless system integration. 
However, it introduces significant control overhead with a bulky host CPU, i.e. a 5-stage pipeline in-order 64-bit Rocket control core. The CPU host occupies 
$1.47 \times$ more area than the spatial GeMM array~\cite{genc2021gemmini}.
2.) It is hard to maintain a high utilization across diverse workloads, causing significant penalties on the system performance. 
Acceleration platforms must be carefully designed to minimize spatial and temporal underutilization of the DNN accelerators since each model/layer has different computation characteristics~\cite{park2022conna, 10129330, yi2022nnasim}.
Low spatial utilization arises from inefficient exploitation of the available parallelism while low temporal utilization stems from idle cycles caused by long configuration times or memory stalls~\cite{pouya2023diana}. 

To overcome the challenges towards a highly efficient, yet flexible and reusable DNN acceleration platform, we present \textit{OpenGeMM}, an open-source\footnote{\textit{OpenGeMM} is available open-source at \url{https://github.com/KULeuven-MICAS/snax_cluster}.} and programmable GeMM acceleration platform with high flexibility and efficiency for edge AI. \textit{OpenGeMM}'s overall system architecture is presented in Figure \ref{fig:sys_arch}. Specifically, 

\setitemize[0]{leftmargin=16pt}
\begin{itemize}
\item We propose a highly parameterized Chisel-coded GeMM accelerator hardware generator with $3$D spatial unrollings, offering ease of customization and efficient data reuse (Section $2$).
\item We provide an efficient system integration platform with a lightweight RISC-V host processor and a tightly coupled memory subsystem, named \textit{OpenGeMM}.
For maximizing system utilization, three mechanisms are introduced, namely configuration pre-loading, input pre-fetching with output buffering, and programmable strided memory access (Section $3$).
\item Experimental results show that \textit{OpenGeMM} achieves sustained high utilization ranging from $81.89$\% to $99.34$\% across diverse DNN workloads. Besides, $3.58 \times$ to $16.40 \times$ normalized throughput speedup is derived compared to open-source state-of-the-art (SotA) accelerator solutions stemmed from \textit{OpenGeMM}'s high system utilization. Moreover, \textit{OpenGeMM} shows $4.68$ TOPS/W system power efficiency and the best operation-area efficiency among peer solutions (Section 4). 
\end{itemize}

%% file: content/gemm_arch.tex
\section{Efficiency and Versatility in GeMM Accelerator Generation}
\label{sec:gemm_arch}

We start by diving into the GeMM accelerator hardware generator, which aims for efficiency across a wide variety of workloads, through a combination of design time configurability and run-time programmability, to always ensure maximal spatial and temporal data reuse. 

\subsection{GeMM Accelerator Dataflow}

\begin{figure}[thbp]
    \centering
    \includegraphics[width=.9\columnwidth]{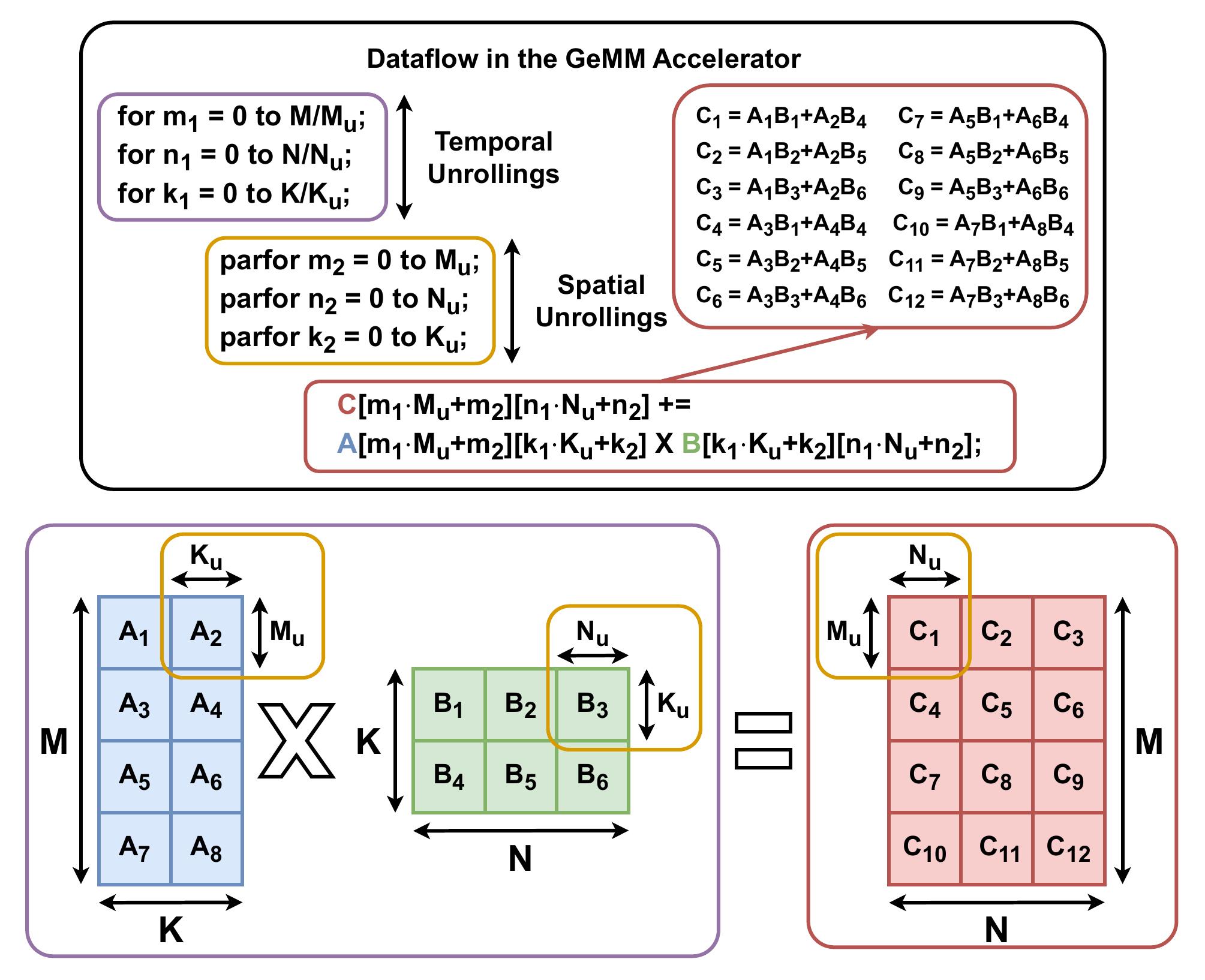}
    \centering
    \caption{Dataflow representation in the GeMM accelerator hardware generator. 
    } 
    \vspace{-0.3cm}
    \label{fig:dataflow}
\end{figure}

The dataflow of the GeMM accelerator is depicted in Figure \ref{fig:dataflow}. The accelerator targets GeMM operations of dimension $(M, K, N)$ as shown in Equation \ref{eq:gemm}:

\begin{equation}
    C_{M,N}=A_{M,K} \times B_{K,N} 
\label{eq:gemm}
\end{equation}

The computation process involves splitting large matrices into tiles which can be executed in one cycle on the GeMM accelerator. Within \textit{OpenGeMM}, this process is represented as 6 nested loops,  which can be further categorized into spatial unrollings and temporal unrollings. Spatial unrollings (the 3 inner-most loops in Figure~\ref{fig:dataflow}) represent spatial computation parallelism within a single clock cycle. 
Temporal unrollings (the 3 outer-most loops in Figure~\ref{fig:dataflow}) reflect the sequential processing order of different tiles, while optimal unrollings will target to maximize local data reuse and minimize the stalls through proper loop ordering.

\subsection{Exploiting 3D Spatial Unrollings to Maximize Spatial Data Reuse}
\label{sec:gemm_datapath_arch}

\begin{figure}[t]
    \centering
    \includegraphics[width=1\columnwidth]{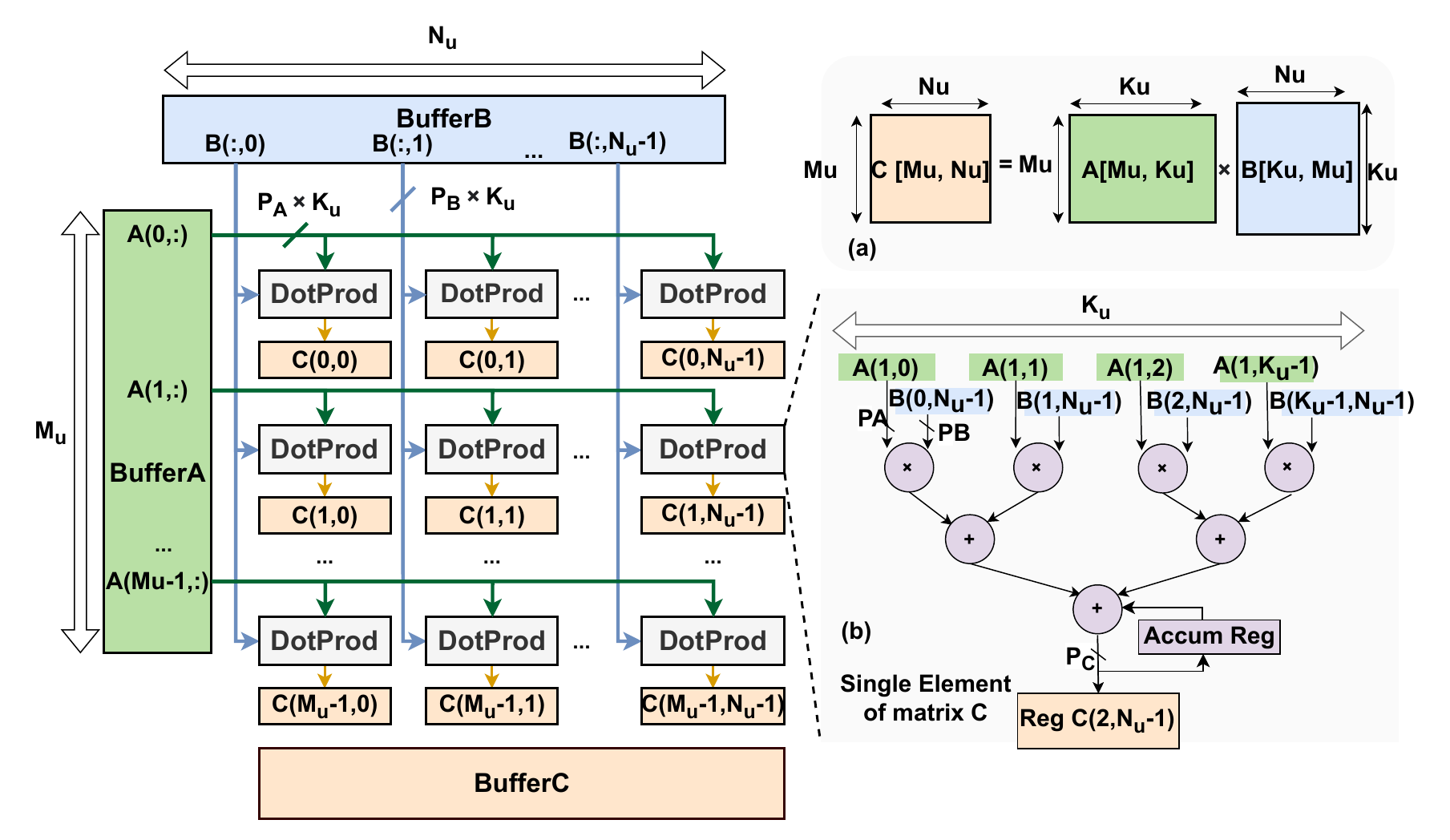}
    \centering
    \caption{GeMM accelerator hardware generator microarchitecture. (a)~Matrix multiplication that GeMM accelerator processes in one cycle with 3D spatial unrollings. (b)~DotProd microarchitecture.
    }
    \vspace{-0.6cm}
    \label{fig:gemm_microarch}
\end{figure}

GeMM accelerator spatially process a tile matrix $A'$ of size $(M_u, K_u)$ and a tile matrix $B'$ of size $(K_u, N_u)$ to produce a tile matrix $C'$ of size $(M_u, N_u)$, as shown in Figure \ref{fig:gemm_microarch}(a).
To maximize spatial data reuse, it is important to reuse every fetched data element of $A'$ and $B'$ as much as possible.
To ensure this, the GeMM array datapath is conceptualized as a 3D MAC (Multiply-Accumulate) array, as depicted in Figure \ref{fig:gemm_microarch}.  The 3D MAC array is organized as a $(M_u, N_u)$-sized mesh of $K_u$-sized vector dot product units (\textit{DotProd} as detailed in Figure \ref{fig:gemm_microarch}(b)) to spatially unroll all dimensions of matrices $A'$, $B'$, and $C'$. 
The 3D MAC array is adapted to match with the 3 nested spatial unrollings for GeMM processing.
Specifically, vectors from $A'$ matrix and vectors from $B'$ matrix are broadcasted horizontally and vertically among the DotProd array, maximizing spatial data reuse.
Within one DotProd, $K_{u}$ multiplication results are combinatorially accumulated to get one result of $C'$.

\textit{OpenGeMM} supports design-time configurations of the DotProd array size and the size of each DotProd unit. This feature enables a flexible
generation of a wide range of accelerators, such as dot-product units, outer-dot product units, vector-matrix multiplication accelerators, or matrix-matrix multiplication accelerators. 
Therefore, varied optimized spatial unrollings can be implemented based on these $(M_u, K_u, N_u)$ parameters to accommodate the diverse computational requirements issued from different applications.

\subsection{Exploiting Temporal Data Reuse to Minimize Memory Access}

Temporal unrollings should be properly ordered to maximize the local data reuse and avoid unnecessary memory conflicts. Depending on whether the innermost temporal loop is weight-relevant or output-relevant, the dataflow can typically be categorized into weight-stationary and output-stationary.
While weight-stationery keeps weight unchanged, output stationary dataflow fits better with GeMM operation, which can be verified through existing DNN dataflow design space exploration (DSE) frameworks~\cite{mei2021zigzag}. The underlying rationale is the precision of the partial sum is often larger than the weight, leading to higher cost when the partial sum is to be updated every cycle.
Take a typical convolutional layer with input tensor size of $(Ox, Oy, C)$ and kernel size of $(K, Fx, Fy, C)$ as an example, after im2col~\cite{anderson2017low}, the convolution operation is translated to matrix multiplication with matrix A of the size of $(O_x \cdot O_y, F_x \cdot F_y \cdot C)$ and matrix B of the size of $(F_x \cdot F_y \cdot C, K)$. Since $F_x \cdot F_y \cdot C$ is typically much larger than $O_x \cdot O_y$, reusing the partial sum in $F_x \cdot F_y \cdot C$ dimension temporally saves more data bandwidth and energy. With this observation, \textit{OpenGeMM} implements an output stationary dataflow supported by an output accumulation register inside each DotProd unit, with the innermost temporal loop iterating from $k1=0$ to $K/Ku$.

All these 6 nested loop processes are handled through a built-in hardware loop controller within the GeMM accelerator, which is in charge of the timely input data request, outputting of result data, and accumulator resets. 
GeMM accelerator can be programmed at run-time with maximum hardware loop upper bound when the required data amount reaches the on-chip buffer capacity.  For even larger matrices, the GeMM accelerator can be called multiple times through software controllers, eg., RISC-V core, to handle extra tiling as more nested temporal loops on higher-level memories.

%% file: content/system_arch.tex
\section{System architecture for high utilization and programmability}
\label{sec:sys_arch}

\begin{figure*}[t]
    \vspace{-0.9cm}
    \centering
    \includegraphics[width=2\columnwidth]{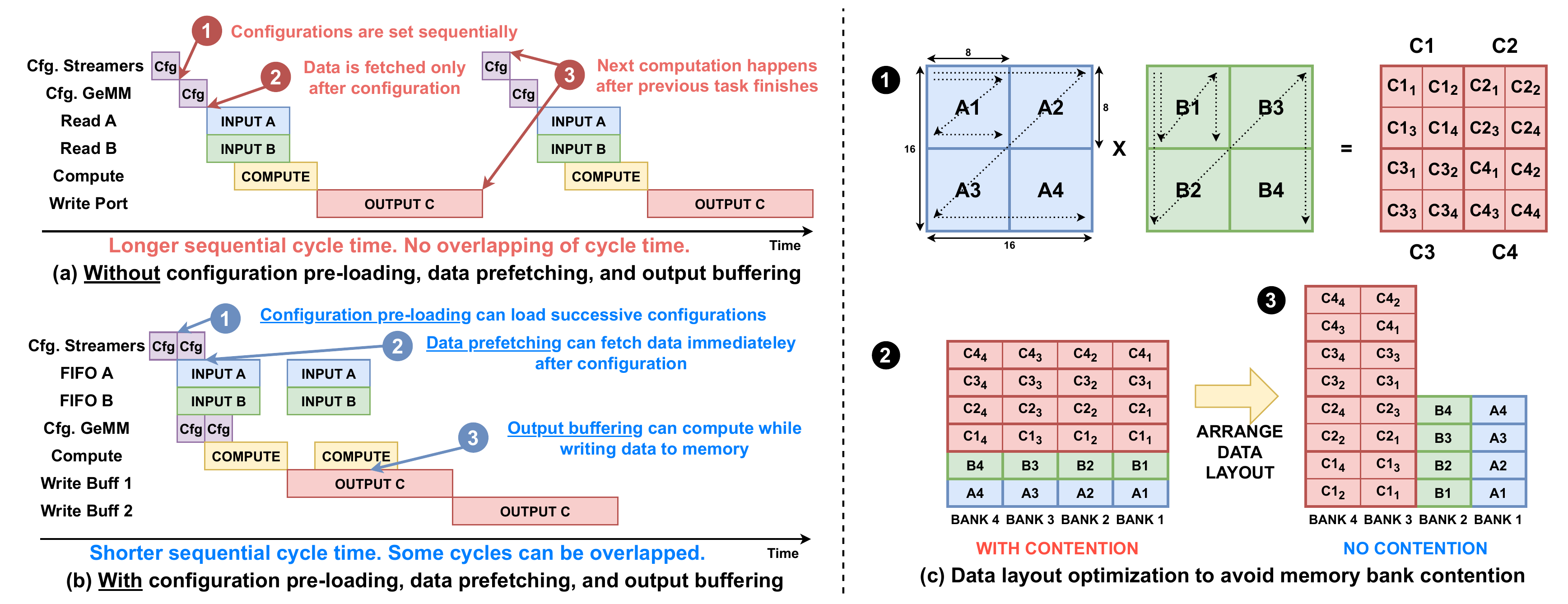}
    \vspace{-0.3cm}
    \centering
    \caption{ (a) Scenario without configuration pre-loading, input pre-fetch, and output buffering. (b) Conceptual visualizations for configuration pre-loading, input pre-fetch, and output buffering. (c) Data layout optimization example. 
    }
    \label{fig:utilization_figs}
    \vspace{-0.3cm}
\end{figure*}

\subsection{OpenGeMM System Architecture}

To guarantee the GeMM accelerator can be smoothly integrated with standard SoCs,
we propose the \textit{OpenGeMM} platform (Figure~\ref{fig:sys_arch}) to enhance programmability and maximize GeMM core throughput. 
It incorporates the GeMM accelerator mentioned earlier, as well as a compact RV32I host core, a tightly coupled multi-bank scratchpad memory, and three data streamers for efficient memory access. \textit{OpenGeMM} offers extensive customization capabilities, with design time parameters summarized in Table \ref{tab:gemm_para}.

\begin{table}[h]
\centering
\begin{threeparttable}
\caption{Customizable \textit{OpenGeMM} design-time parameters and case study configuration.}
\label{tab:gemm_para}
\begin{tabular}{ccc}
    \toprule
    Parameters & Meaning  & Case study values\\
    \hline
    \multicolumn{3}{c}{\textbf{Parameters for GeMM core}} \\
    \hline
    $M_{u}$ &Number of rows of the array & 8\\
      $N_{u}$ & Number of columns of the array& 8\\
     $K_{u}$ & Size of each DotProd & 8\\
    \hline 
     $P_A$ & Integer bit precision of $A$ & 8\\
       $P_B$ & Integer bit precision of $B$& 8\\
       $P_C$ & Integer bit precision of $C$ & 32\\
    \hline
    \multicolumn{3}{c}{\textbf{Parameters for memory system}} \\
     \hline
   \multirow{2}*{$D_{stream}$} & Pre-fetch buffer & \multirow{2}*{3} \\
   & and output buffer depth  & \\
   $R_{mem}$ & Input memory ports  & 16\\
   $W_{mem}$ & Output memory ports & 32\\
    $P_{word}$ & Memory port data width & 64 \\
    $N_{bank}$ & Number of banks& 32\\
    $D_{mem}$ & Bank depth  & 1056 \\
        \bottomrule
  \end{tabular}
\end{threeparttable}
\vspace{-0.3cm}
\end{table}

Specifically, \textit{OpenGeMM} utilizes a lightweight 32-bit integer RISC-V Snitch core \cite{zaruba2020snitch} for programming the GeMM compute core using standard RISC-V Configuration and Status Registers (CSR) instructions \cite{riscv-priv-arch}, while minimizing control area overhead.
A specific register address range is allocated for the accelerator CSR configurations. 
A dedicated CSRManager module facilitates communication between the Snitch core and GeMM core through CSR operations, achieving high configuration bandwidth (32 bits/cycle). Multiple accelerator configurations can be consolidated into a single CSR to optimize configuration cycles.
Since the CSR instruction is already part of the RISC-V ISA, there is no need to modify the RISC-V control core and the compiler, greatly alleviating the accelerator system integration and programming challenge.

To ensure efficient memory access, we tightly couple a wide-bandwidth, software-controlled multi-banked scratchpad memory (SPM) to the GeMM core. 
The multi-banked SPM supports configurable number of read ($R_{mem}$) and write ($W_{mem}$) ports with data width ($P_{word}$), organized into multiple banks ($N_{bank}$). Additionally, data streamers between the multi-banked SPM and GeMM core employ programmable hardware loops for efficient and autonomous data access catering to the GeMM core in a streaming way.

To improve the utilization problem in existing works,
three system-level mechanisms are further introduced within \textit{OpenGeMM},
which are detailed in subsequent subsections.

\subsection{Configuration Pre-loading} 

Due to the sequential programming of numerous CSRs, including hardware loop bounds related to workload size $(M, K, N)$, base addresses and strides related to memory access, the programming cycle can be lengthy, as shown in Figure \ref{fig:utilization_figs}(a) \idr{1}. To hide the configuration time, we introduce a configuration pre-loading (CPL) mechanism. It allows the host to pre-load the GeMM core configuration for the next computation, while the current computation is being executed, effectively overlapping configuration time with computation time, as depicted in Figure \ref{fig:utilization_figs}(b) \idblue{1}.

\subsection{Input Pre-fetch and Output Data Buffering}

To minimize computation stalls, we incorporate a data buffer within the data streamer to allow continuous pre-fetching of data until the buffer reaches its capacity. As long as there is space in the buffer (data being consumed by the GeMM core), it will actively pre-fetch data. This dynamic producer-consumer mechanism aims to maintain high utilization of the GeMM array. The buffer depth is a configurable parameter set at design time, based on the amount of data that would like to be pre-fetched.
Figure \ref{fig:utilization_figs}(b) \idblue{2}~ illustrates that immediately after configuring the streamer, the data buffers begin pre-fetching data while the GeMM core is being configured.

The output stationary nature ensures that the GeMM core writes back a result every $K/K_{u}$ computational cycles, facilitating the reduction of output memory stalls through an output buffering mechanism. Specifically, a configurable number of output data buffers alternate between storing the output from the GeMM core and writing the output matrix to memory in a round-robin fashion.
Figure \ref{fig:utilization_figs}(b) \idblue{3}~ illustrates the timing utilization of the write ports. This allows computation to proceed concurrently with writing the output of the last result.
Without the input pre-fetch and output data buffering, the GeMM core utilization can be lower because of idle cycles caused by input and output memory stalls, as shown in Figure \ref{fig:utilization_figs}(a) \idr{2}~ and \idr{3}.

\vspace{-0.2cm}
\subsection{Strided Memory Access}

The compiler can exploit strided data access \cite{schuiki2020stream} to optimize the data layout by interleaving access, thereby minimizing bank conflicts and maximizing the utilization of the GeMM core.
Figure \ref{fig:utilization_figs}(c) illustrates an example of data layout optimization. Figure \ref{fig:utilization_figs}(c) \idb{1}~ shows a set of matrices and their sub-matrices for GeMM operation input. These matrices can be organized contiguously in either row-major or column-major order in memory as depicted in Figure \ref{fig:utilization_figs}(c) \idb{2}. This layout will lead to bank contentions because sub-matrices A1 and B1 are accessed in the same bank. However, by transforming the data layout as shown in Figure \ref{fig:utilization_figs}(c) \idb{3}, bank contentions can be avoided.

To address this, we equip each data streamer with a configurable strided address generation \cite{schuiki2020stream} unit (AGU) to support flexible data access to relieve the memory bank contentions. Specifically,

\begin{itemize}
    \item \textit{At design time,} we configure the AGU such that it matches the GeMM core's port width and the address generation pattern, such as how many nested loops are needed for strided address generation, of operands $A$, $B$, and $C$. These are also related to the number of multi-banked SPM read ports $R_{mem}$, write ports $W_{mem}$, and their width $P_{word}$.
    \item \textit{At run time,} we program the hardware loop bounds, base addresses, and two-dimensional memory strides for each data streamer to generate the corresponding data layout address. 
\end{itemize}

%% file: content/results.tex
\section{OpenGeMM evaluation and SotA comparison}

\subsection{Evaluation Setup}

We evaluate the proposed \textit{OpenGeMM} system at the register transfer level (RTL), utilizing Chisel \cite{bachrach2012chisel} for designing the GeMM accelerator hardware generator and SystemVerilog to implement all other platform components built upon~\cite{zaruba2020snitch}. We generate one \textit{OpenGeMM} instance using the case study parameters listed in Table \ref{tab:gemm_para}. We choose an 8x8x8 GeMM array to achieve a good balance between spatial utilization and hardware throughput for typical GeMM workloads. 
For performance and utilization evaluation, we conduct cycle-accurate RTL simulation using Verilator. 
The system is then synthesized with Synopsys Design Compiler under the TSMC 16nm FFC process technology at a clock frequency of 200MHz and a supply voltage of 0.675V. Power analysis is conducted using Synopsys PrimeTime.

\subsection{Utilization Analysis}
An ablation experiment is performed to assess the effectiveness of the proposed three mechanisms in enhancing the GeMM core's utilization. 
We conduct experiments on the 8x8x8 GeMM accelerator with 500 different computational matrix sizes $(M, K, N)$, randomly choosing from $M, K, N \in \{8, 16, 24, \ldots, 256\}$.
To access the benefits of proposed mechanisms, the combinations consisting of different techniques are enabled to evaluate their individual impact on utilization. Each workload is repeated 10 times to observe the effect of the configuration pre-loading mechanism.

\begin{figure}[thbp]
    \centering
    \vspace{-0.3cm}
    \includegraphics[width=0.47\textwidth]{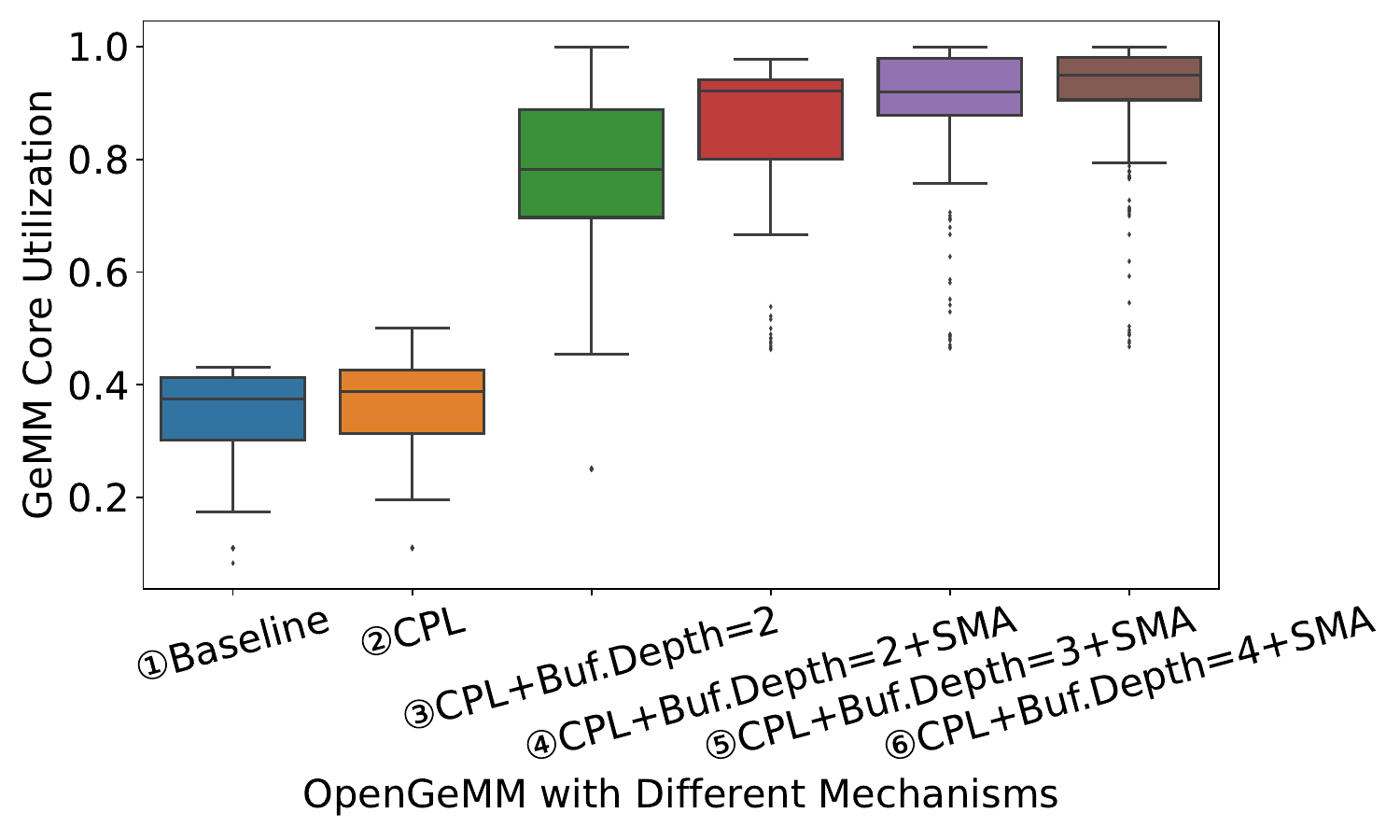}
    \vspace{-0.4cm}
    \centering
    \caption{Utilization analysis of \textit{OpenGeMM} under 500 random computational matrix sizes and different utilization enhancement techniques. 
    }
    \vspace{-0.5cm}
    \label{fig:utilization}
\end{figure}

Figure~\ref{fig:utilization} hence presents the utilization results in a box-plot format. Specifically, incorporating the configuration pre-loading (CPL) (Arch\ding{173}) results in a median utilization improvement of $1.4\times$ compared with baseline (Arch\ding{172}). 
When combining input pre-fetching and output buffering (Buf.Depth=2) (Arch\ding{174}), the median utilization improves by an additional $2.02\times$ compared with Arch\ding{173}, demonstrating the effectiveness of the input data pre-infecting and output data buffering. Additionally, employing strided memory access (SMA) (Arch\ding{175}) increases median utilization by $1.18\times$ compared with Arch\ding{174}, showing the data layout optimization effect. 
Overall, implementing all three techniques (Arch\ding{175}) enhances median utilization by $2.78\times$ compared to the baseline \textit{OpenGeMM} platform (Arch\ding{172}). Exploring the impact of increasing buffer depth to 3 and 4 shows a consistent increase in utilization with less variation across different matrix sizes.


\begin{table}[h]
  \centering
  \vspace{-0.1cm}
  \begin{threeparttable}
  \caption{Utilization (in \%) and performance (in cycles) of \textit{OpenGeMM} on real DNN workloads.} 
  \label{tab:ai_workload} 
    \begin{tabular}{lcccc}
        \toprule
        & \textbf{MobileNetV2} & \textbf{ResNet18} & \textbf{ViT-B-16} & \textbf{BERT-Base} \\
        \midrule
        \textbf{SU}\tnote{$\ast$} & 87.36 & 96.01 & 98.41 & 99.54 \\
        \textbf{TU}\tnote{$\dagger$} & 93.74 & 99.72 & 99.75 & 99.80 \\
        \textbf{OU}\tnote{$\ddagger$} & 81.89 & 95.74 & 98.16 & 99.34 \\
        \textbf{CC}\tnote{$\S$} & \(3.33\times 10^8\) & \(9.29\times 10^8\) & \(1.79\times 10^{10}\) & \(4.93\times 10^{10}\) \\
        \bottomrule
    \end{tabular}
    \begin{tablenotes}
    \footnotesize
    \item \begin{minipage}[t]{0.48\linewidth}[$\ast$] Spatial utilization.\end{minipage}%
    \begin{minipage}[t]{0.48\linewidth}[$\dagger$] Temporal utilization.\end{minipage}
    \item \begin{minipage}[t]{0.48\linewidth}[$\ddagger$] Overall utilization.\end{minipage}%
    \begin{minipage}[t]{0.48\linewidth}[$\S$] Cycle count.\end{minipage}
    \end{tablenotes}
     \end{threeparttable}
    \vspace{-0.5cm}
\end{table}

\begin{table*}[htb]
  \vspace{-0.8cm}
  \centering
  \normalsize
  \begin{threeparttable}
  \caption{State-of-the-Art Comparison. The efficiency metrics shown for each platform are system efficiencies.
  }   
  \vspace{-0.4cm}
  \label{tab:sota_compare} 
      \begin{tabular}{c|c|c|c|c|c|c|c}
        \toprule
             \multirow{2}*{\textbf{Accelerator}} &  \textbf{SIGMA} & \textbf{CONNA} & \textbf{Gemmini} & \textbf{DIANA} & \textbf{RBE} & \textbf{RedMule} & \textbf{\textit{OpenGeMM}} \\
             & \cite{qin2020sigma} & \cite{park2022conna} & \cite{genc2021gemmini} & \cite{pouya2023diana}  & \cite{conti2024marsellus} & \cite{tortorella2023redmule}  & This Work\\
        \midrule
            \textbf{Tech ($\textbf{nm}$)} & 28 & 65 & 22 & 22 & 22 & 22 & 16 \\
            \textbf{Area ($\textbf{mm2}$)} & 65 & 2.36 & 1.03 & 8.91 & 2.42 & 0.73 & 0.62\tnote{\textdagger} \\
            \textbf{Memory (KiB)} & 6,000 & 144 & 256 & 512 & 128 & 128 & 270 \\
            \textbf{Freq (MHz)} & 500 & 200 & 1000 & 280 & 420 & 470 & 200 \\
            \hline
            \multirow{2}*{\textbf{Peak Perf. (GOPS)}} & \multirow{2}*{16,000} & \multirow{2}*{102.4} & \multirow{2}*{512} & 224 (Dig.) & 637 (2b) & \multirow{2}*{89} & \multirow{2}*{204(8b)} \\
            & & & & 40 (AIMC) & 91 (8b) & & \\
            \multirow{2}*{\textbf{Peak Eff. (TOPS/W)}} & \multirow{2}*{0.48} & \multirow{2}*{0.856} & \multirow{2}*{-} & 1.7 (Dig.) & 12.4 (2b) & \multirow{2}*{1.6} & \multirow{2}*{4.68(8b)} \\
            & & & & 4 (AIMC) & 0.74 (8b) & & \\
            \multirow{2}*{\textbf{Peak Perf./Area (GOPS/mm2)}} & \multirow{2}*{246} & \multirow{2}*{43} & \multirow{2}*{497} & 25 (Dig.) &  263 (2b) & \multirow{2}*{121} & \multirow{2}*{329(8b)}\tnote{\textdagger}  \\
            & & & & 4.5 (AIMC) & 37 (8b) & & \\
            \multirow{2}*{\textbf{Op-Area-Eff. (TOPS/W/mm2)}} & \multirow{2}*{0.0073} & \multirow{2}*{0.363} & \multirow{2}*{-} & 0.2 (Dig.) &  5.12 (2b) & \multirow{2}*{2.2} & \multirow{2}*{7.55(8b)}\tnote{\textdagger}  \\
            & & & & 0.44 (AIMC) & 0.31 (8b) & & \\
            \textbf{Supported Precision} & BFP 16, FP 32 & INT 4, 8, 16, 32 & INT8 & INT 8 & INT 2, 4, 8 & FP 8, 16 & INT 2, 4, 8\tnote{$\S$}\\
            \hline
            \textbf{Open Source} & \checkmark & $\times$ & \checkmark & \checkmark & \checkmark & \checkmark & \checkmark \\
            \textbf{Generated Arch.}\tnote{*} & $\times$ & \checkmark & \checkmark & $\times$ & $\times$ & $\times$ & \checkmark \\
            \textbf{Design- and Run- time Config.}\tnote{$\P$} & $\times$ & \checkmark & \checkmark & $\times$ & $\checkmark$ & $\checkmark$ & \checkmark \\
        \bottomrule
      \end{tabular}
        \begin{tablenotes}
        \footnotesize
          \item[*] Means the design comes from a hardware generator like Chisel. 
          \item[$\P$] Has parameters configurable during design-time (or hardware design configurations) and run-time (or programmable).
          \item[$\dagger$] After placement and routing layout area estimation with 60\% cell density according to \cite{paulin2022soft}.
          \item[$\S$] Design-time configurable.
        \end{tablenotes}
     \end{threeparttable}
    \vspace{-0.4cm}
\end{table*}

\subsection{Real DNNs Benchmarking}

To evaluate the performance of \textit{OpenGeMM} in practical scenarios, several typical CNN and Transformer workloads, including ResNet18 \cite{he2016identity}, MobileNetV2 \cite{sandler2018mobilenetv2}, Vision Transformer \cite{dosovitskiy2020image} and Bert-Base \cite{devlin2018bert} are benchmarked. Our focus is specifically on the energy- and latency-dominant blocks, including convolutional layers (executed via im2col~\cite{anderson2017low}), multi-head attention, multilayer perceptron layers, and fully connected layers within these models.

The utilization and performance\footnote{The data movement cycles between the off-chip DRAM and the on-chip SRAM are not counted.} results of these models are presented in Table \ref{tab:ai_workload}.
Due to the irregular matrix sizes inherent in these workloads, \textit{OpenGeMM} does not achieve full spatial utilization. For example, in the MobileNetV2 model that has rich depth-wise convolutions, the spatial utilization is only $87.36\%$, indicating that the im2col-ed matrix sizes are not multiples of $(M_u, K_u, N_u)$. Moreover, the depth-wise convolutions have fewer channels (tick channels), resulting in smaller K values and slightly lower temporal utilization compared to other workloads. Conversely, in ResNet18 where channel sizes are larger, temporal utilization is as high as $95.74\%$.
In Transformer workloads with regular matrix sizes and no small K values, \textit{OpenGeMM} consistently achieves near $100\%$ spatial and temporal utilization, approaching peak performance.
Despite varying computation characteristics across different workloads, the overall utilization of the GeMM core remains consistently high, ranging from $81.89\%$ to $99.34\%$. This indicates efficient execution of diverse DNN workloads on the run-time programmable \textit{OpenGeMM} platform. 

\begin{figure}[thbp]
    \vspace{-0.1cm}
    \centering
    \includegraphics[width=1.02\columnwidth]{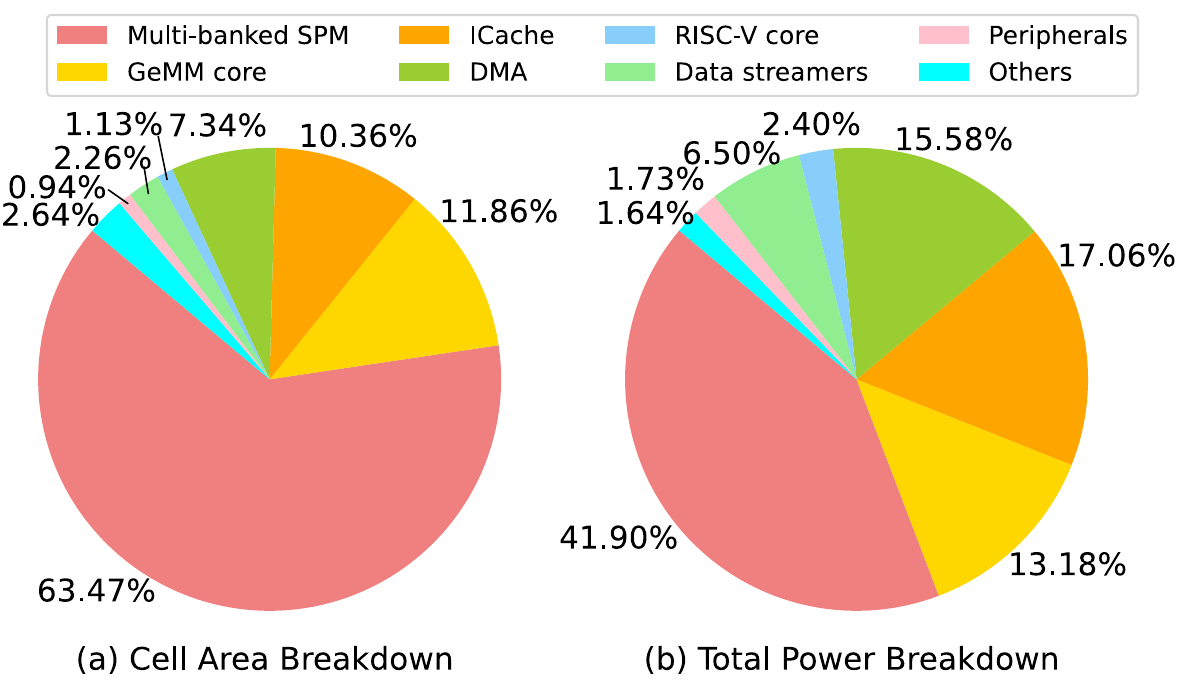}
    \centering
    \caption{\textit{OpenGeMM} cell area and total power breakdown.}
     \vspace{-0.3cm}
    \label{fig:area_power_breakdwon}
\end{figure}

\vspace{-0.1cm}
\subsection{Area and Power Evaluation}
The workload for system power estimation involves block matrix multiplication with a size of $(32, 32, 32)$. 
\textit{OpenGeMM} system occupies a cell area of $0.531~mm^2$ and consumes a total system power of $43.8$ mW when operating at 200MHz. Achieving a peak performance of $204.8$ GOPS, the \textit{OpenGeMM} demonstrates a system efficiency of $4.68$ TOPS/W.

The detailed area and power breakdown of the \textit{OpenGeMM} system is illustrated in Figure \ref{fig:area_power_breakdwon}, with all other system components encompassed, such as the instruction cache and DMA. The largest area component is the $270$KiB multi-banked SPM including the interconnect towards the streamers, occupying $63.47\%$ of the total area breakdown, followed by the GeMM core at $11.86\%$. In terms of power breakdown, multi-banked SPM accounts for $41.90\%$, instruction cache for $17.06\%$, and GeMM core for $13.18\%$. Data streamers for data movement occupy $2.26\%$ of total area and $6.5\%$ of total system power. It is important to note that the RISC-V control overhead is negligible, around $1.13\%$ of the entire system cost and $2.4\%$ of system power, offering \textit{OpenGeMM} efficiency with minimal hardware overheads.

\subsection{State of The Art Comparison} 

We now compare \textit{OpenGeMM} against the state-of-the-art (SotA) of flexible DNN acceleration systems, as summarized in Table \ref{tab:sota_compare}. 
Specifically, we benchmark the throughput of \textit{OpenGeMM} against the SotA GeMM accelerator generation system Gemmini~\cite{genc2021gemmini}, utilizing performance data from \cite{gonzalez202116mm}.

\begin{figure}[t]
    \centering
    \includegraphics[width=0.46\textwidth]{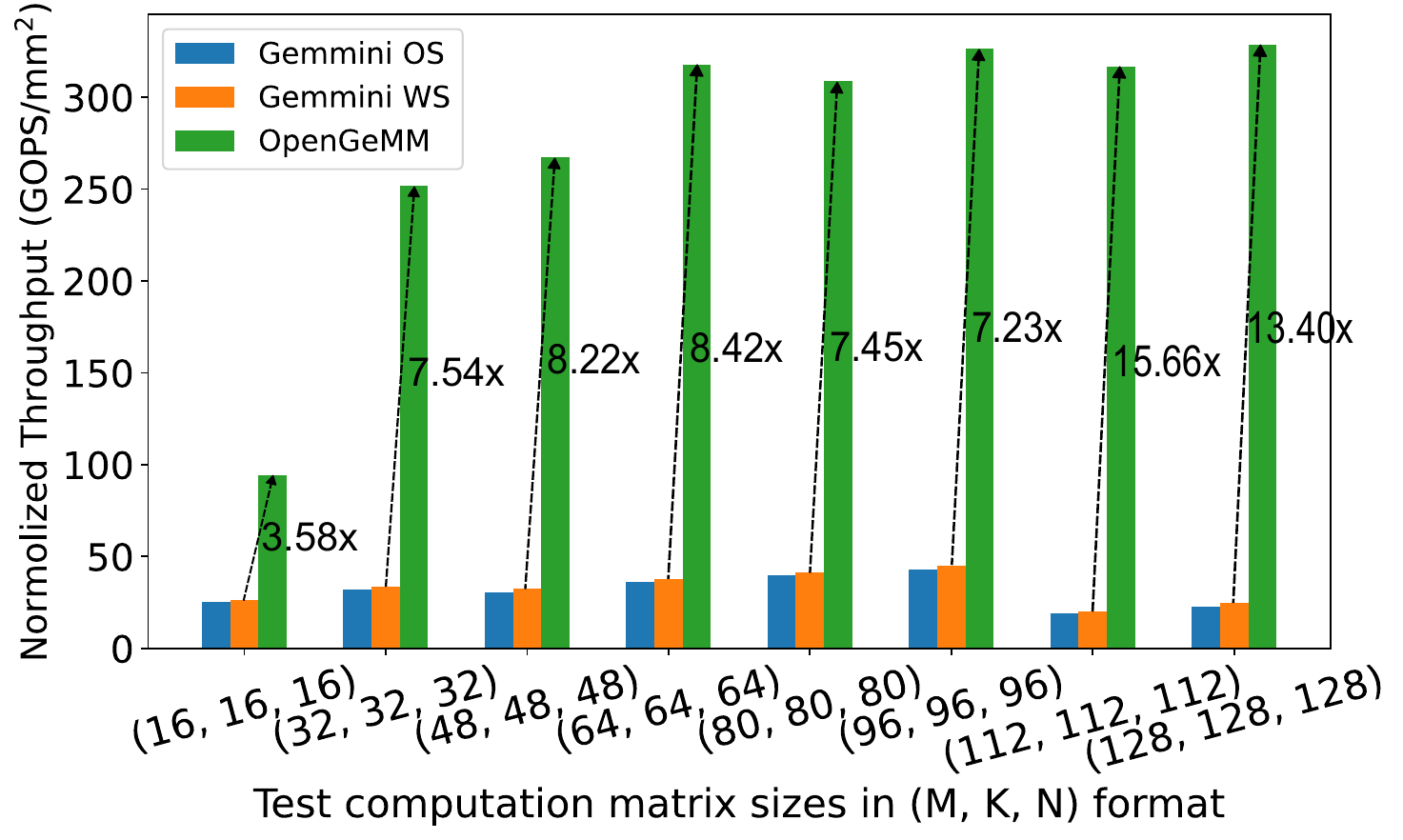}
    \centering
    \caption{Comparison of normilized throughput (GOPS/$mm^2$) for Gemmini~\cite{genc2021gemmini} in output stationary (OS) mode, Gemmini~\cite{genc2021gemmini} in weight stationary (WS) mode, and \textit{OpenGeMM} across various computational matrix sizes.}
    \label{fig:perf_with_gemmini}
    \vspace{-0.5cm}
\end{figure}

Figure \ref{fig:perf_with_gemmini} illustrates the area-normalized throughput comparison (in GOPS/$mm^2$) among the Gemmini in output-stationary (OS) mode and weight-stationary (WS) mode, and \textit{OpenGeMM} across matrix sizes ranging from (8, 8, 8) to (128, 128, 128). 
Compared to Gemmini OS and WS, \textit{OpenGeMM} exhibits normalized throughput speedups ranging from $3.75\times$ to $16.40\times$ and $3.58\times$ to $15.66\times$, respectively.
The exceptional performance of \textit{OpenGeMM} is attributed to the high utilization of the GeMM core, approaching ideal peak performance for these workloads. In contrast, Gemmini falls short of maintaining high temporal utilization (on average $6.25\%$)  on these workloads, because of intensive memory stalls, resulting in lower actual throughput. 

In summary, \textit{OpenGeMM}'s lightweight RISC-V core and tight memory coupling reduce hardware overhead from control and data access while the GeMM core architecture maximizes spatial and temporal data reuse. Moreover, \textit{OpenGeMM} introduces three mechanisms to achieve sustained high utilization. This results in \textit{OpenGeMM} achieving high system energy efficiency and the best operation-area efficiency across all SotAs with int8 data precision.
These features make \textit{OpenGeMM} suitable for efficient edge DNN computing.

%% file: content/conclusion.tex
\section{conclusion}
This paper introduces \textit{OpenGeMM}, an open-source GeMM acceleration platform targeting edge AI applications.
\textit{OpenGeMM} is built around a Chisel-based GeMM accelerator, associated with a lightweight RISC-V processor and a tightly coupled memory system. 
To improve the hardware utilization and therefore the processing efficiency, three mechanisms are introduced at the system level.
The experiments demonstrate that these mechanisms can improve the MAC array utilization of \textit{OpenGeMM} consistently up to $81.89\%$-$99.34$\% across various DNN workloads. Compared with the SotA, \textit{OpenGeMM} shows $4.68$ TOPS/W system power efficiency and $3.58 \times$ to $16.40 \times$ normalized throughput speedup. 

\section*{Acknowledgment}
This project has been partly funded by the European Research Council (ERC) under grant agreement No. 101088865, the European Union’s Horizon 2020 program (CONVOLVE) under grant agreement No. 101070374, the Flanders AI Research Program, Research Foundation-Flanders (FWO) under grant 1SE7723N, and KU Leuven. Guilherme Paim expresses gratitude to the CAPES Brazilian Foundation for their financial support in past grants and for accepting the novation agreement that allowed him to work abroad.